\documentclass[12pt]{iopart}

\usepackage{graphicx}
\usepackage{bm}
\usepackage{epstopdf}
\begin{document}

\title[]{From collective oscillation to chimera state in a nonlocally excitable system}

\author{Qionglin Dai$^1$, Mengya Zhang$^1$, Hongyan Cheng$^1$, Haihong Li$^1$, Fagen Xie$^2$, and Junzhong Yang$^1$}

\address{$^1$School of Science, Beijing University of Posts and
Telecommunications, Beijing, 100876, People's Republic of China\\
$^2$Department of Research and Evaluation, Kaiser Permanente Southern California, Pasadena, CA 91101, USA}

\ead{jzyang@bupt.edu.cn, xiefagen@yahoo.com}
\vspace{10pt}

\begin{abstract}
Chimera states, which consist of coexisting domains of spatially
coherent and incoherent dynamics, have been widely found in
nonlocally coupled oscillatory systems. We demonstrate for the
first time that chimera states can emerge from excitable systems
under nonlocal coupling in which isolated units only allow for the
equilibrium. We theoretically reveal that nonlocal coupling
induced collective oscillation is behind the occurrence of the
chimera states. We find two different types of chimera states,
phase-chimera state and excitability-chimera states, depending on
the coupling strength. At weak coupling strength where collective
oscillation is localized around the unstable homogeneous
equilibrium, the chimera states are similar to the ones in
nonlocally coupled phase oscillators. For the chimera states at
strong coupling strength, the dynamics of both coherent units and
incoherent units shift back and forth between low amplitude
oscillation induced by collective oscillation and high amplitude
oscillation induced by excitability of local units.
\end{abstract}

%
\vspace{2pc}

\section{Introduction}
Chimera states refer to a type of fascinating hybrid
dynamical states in which identically coupled units spontaneously
develop into coexisting synchronous and asynchronous parts. The
chimera states were first found numerically by Kuramoto and his
colleagues in 2002 \cite{kura}, theoretically investigated by
Strogatz and his colleagues in 2004 \cite{abra}, and then  have
become a very active research field \cite{pan15,abr08,pik08}. The
chimera states were realized experimentally in
chemical \cite{tins12,sch14}, optical \cite{hage12,vik14},
electronical \cite{larg13}, mechanical and electrochemical
systems \cite{mart13,kap14,oml15,wick13}. Chimera states are also
possibly related to Parkinson's diseases where synchronized
activities are absent in certain regions of the brain \cite{levy00}, and epileptic seizures where specific regions of
the brain are highly synchronized while other parts are not
synchronized \cite{mot10,aya73}. Different types of chimera states
such as breathing chimeras \cite{abra}, amplitude-mediated
chimeras \cite{zak14}, multi-cluster
chimeras \cite{omel14,mais14,zhu12}, and spiral
chimeras \cite{mart10a,gu13} have been discovered and investigated
in details.

Chimera states are initially observed in nonlocally coupled phase
oscillators, and both nonlocal coupling and phase oscillators are
thought to be required for realizing chimera states. But it has
been extensively observed that chimera states can occur in
globally coupled \cite{set14,sch15} and locally coupled
oscillators \cite{lai15}, periodic and chaotic maps \cite{omel11},
Stuart-Landau models \cite{lai10,zak14}, Van der Pol
oscillators \cite{omel15}, FitzHugh-Nagumo (FHN)
oscillators \cite{omel13}, Hindmarsh-Rose models \cite{hiz14},
Hodgkin-Huxley models \cite{sak06} and delayed-feedback
systems \cite{larg13}. However, these systems still have one common
requirement that each individual unit is self-oscillating when it
is isolated. Then one great and interesting question arises: if
isolated unites do not support self-oscillation, can chimera
states be realized when they are nonlocally coupled?

Excitable systems with a stable equilibrium are ubiquitous in
biology, chemistry, and physics \cite{win72,zyk87}. In an
excitable system, it takes a large excursion before going back to
the equilibrium if perturbations are stronger than a threshold.
Recently, chimera states have been studied in nonlocally coupled
type-I excitable systems \cite{vul14} and nonlocally coupled
excitable systems in the presence of noise \cite{sem16}. But the
isolated units in these works are still self-oscillating rather
than the follow a stable equilibrium. In this paper, we will
demonstrate that chimera states can be realized in a nonlocally
coupled excitable system in which each isolated unit only allows
for an equilibrium rather than a self-oscillation.

The rest of paper is organized as follows. In section 2, we
present the model and numerical demonstrate the existence of
chimera dynamics. In section 3, we present the stability
analysis on the homogeneous equilibrium and demonstrate the
emergence of collective oscillation, followed by the numerical
results on the transition scenario from homogeneous equilibrium to
chimera dynamics. In section 4, we address the roles of model
parameters on chimera dynamics. Finally, we conclude with a
summary in section 5.

\begin{figure*}
\includegraphics[width=6in]{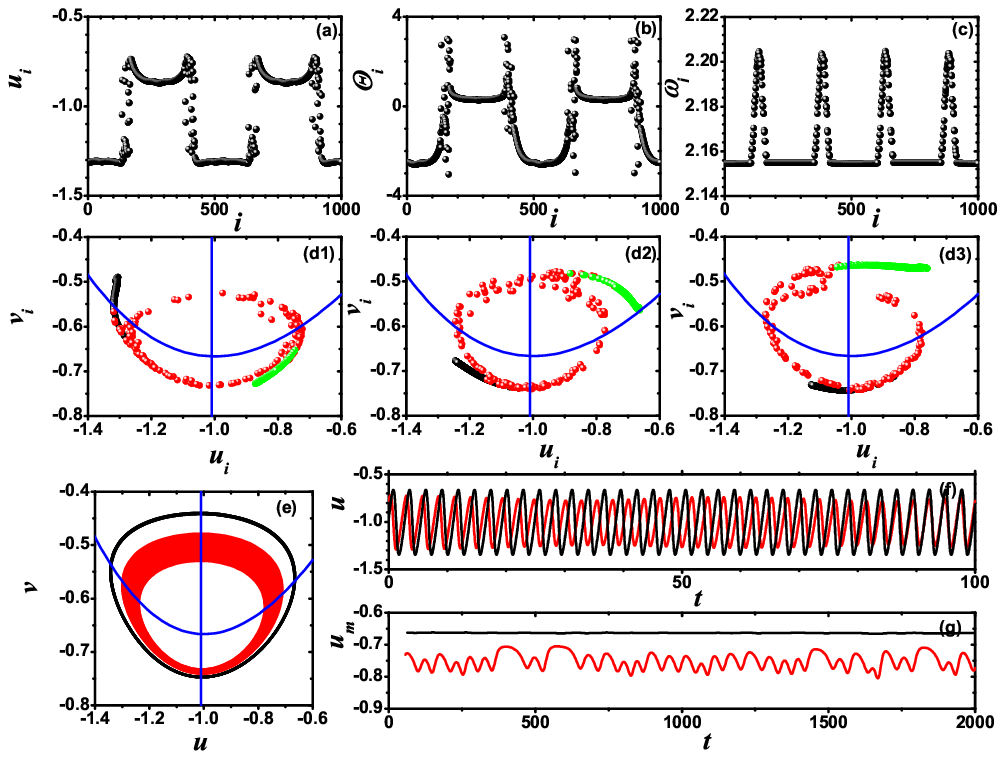}
\caption{\label{fig1}(color online)(a) Snapshot of the variables
$u_i$. (b) Snapshot of the phases $\Theta_i$. (c) Mean phase
velocities $\omega_i$. (d1)-(d3) Snapshots in the ($u_i,v_i$)
plane at different times (blue lines denote the nullclines of the
FHN unit). Incoherent units denoted by red dots, the units in
adjacent clusters denoted by black and green dots, respectively.
(e) The trajectories of one coherent (in black) and one incoherent
(in red) FHN units. (f) The time sequence of the coherent (in
balck) and the incoherent (in red) units. (g) The evolution of the
maximum of $u$, $u_m$ for the coherent (black) and the incoherent
I in red) units. $\phi=\pi/2+0.57$, $\sigma=0.05$. In the
numerical simulations, we employ a fourth-order Runge-Kutta
algorithm with a time step $\delta t=0.01$. Random initial
conditions are used. }
\end{figure*}

\section{The model}
We consider a one-dimensional ring of $N$ nonlocally
coupled FHN systems \cite{omel13} in which the individual unit is
coupled to $R$ neighbors on each side with coupling strength
$\sigma$:
\begin{eqnarray}\label{eq1}
\epsilon\dot{u}_i&=&u_i-\frac{u_i^3}{3}-v_i \nonumber\\
&+&\frac{\sigma}{2R}\Sigma_{j=i-R}^{i+R}[b_{uu}(u_j-u_i)+b_{uv}(v_j-v_i)],\nonumber\\
\dot{v}_i&=&u_i+a\nonumber\\
&+&\frac{\sigma}{2R}\Sigma_{j=i-R}^{i+R}[b_{vu}(u_j-u_i)+b_{vv}(v_j-v_i)].
\end{eqnarray}
The subscript $i$ refers to the unit index, which has to be taken
module $N$ (or period boundary condition). $u_i$ and $v_i$ are the activator and inhibitor
variables, respectively. $\epsilon>0$ is the parameter
characterizing a time separation. Following the reference \cite{omel13}, the coupling matrix is modelled as:
\begin{eqnarray}\label{eq2}
B=\left(
    \begin{array}{cc}
      b_{uu} & b_{uv} \\
      b_{vu} & b_{vv} \\
    \end{array}
  \right)=\left(
    \begin{array}{cc}
      \cos\phi & \sin\phi \\
      -\sin\phi & \cos\phi \\
    \end{array}
  \right).
\end{eqnarray}
It is convenient to consider the ratio $r=R/N$, the coupling
radius, which ranges from $1/N$ (nearest neighbor coupling) to
$0.5$ (globally coupling). Isolated FHN unit exhibits excitable
behavior for $|a|>1$ while oscillatory behavior for $|a|<1$ via a
Hopf bifurcation at $|a|=1$. Because our study focuses on the
excitable regime, we set $a=1.01$, $\epsilon=0.18$, $r=0.35$
and $N=1000$ throughout the letter. So the isolated units of
Eq.(\ref{eq1}) have an equilibrium $(u^*,v^*)=(-a,-a+a^3/3)$.

\begin{figure*}
\includegraphics[width=6in]{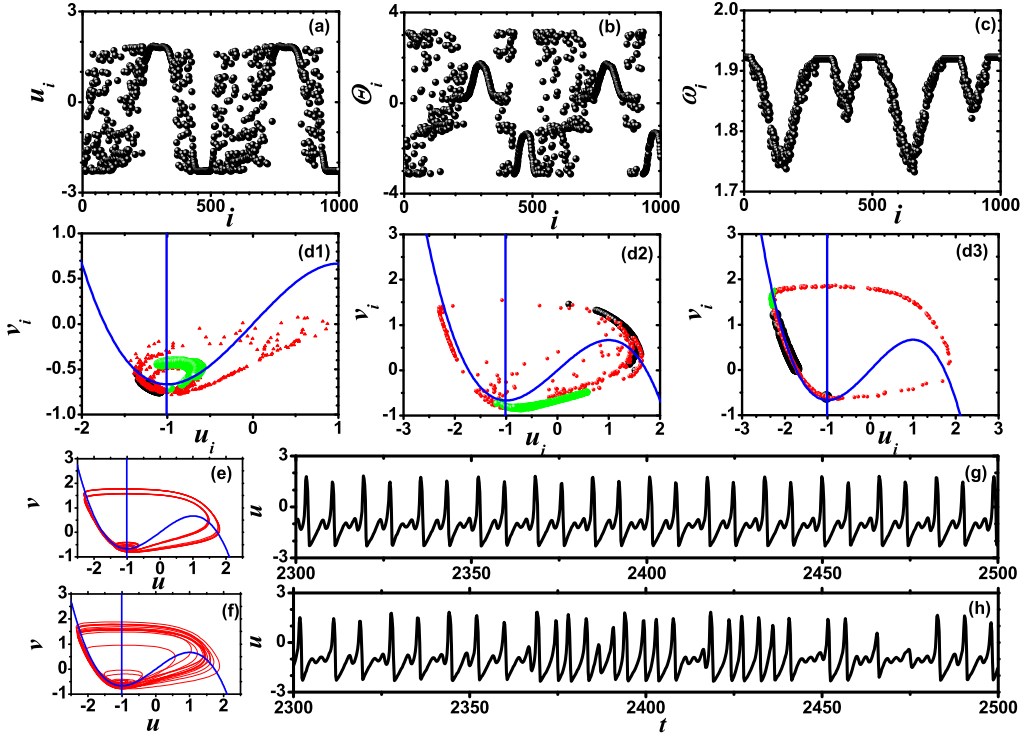}
\caption{\label{fig2}(color online)(a) Snapshot of the variables
$u_i$. (b) Snapshot of the phases $\Theta_i$. (c) Mean phase
velocities $\omega_i$. (d1)-(d3) Snapshots in the ($u_i,v_i$)
plane at different times (blue lines denote the nullclines of the
FHN unit). Incoherent units denoted by red dots, the units in
adjacent clusters denoted by black and green dots, respectively.
(e) and (f) The trajectories of one coherent and one incoherent
FHN units, respectively. (g) and (h) The time sequence of the
coherent and the incoherent units, respectively.
$\phi=\pi/2+0.57$, $\sigma=0.18$.}
\end{figure*}

By increasing the coupling strength $\sigma$, we find that chimera
states do exist for nonlocally coupled excitable FHN units.
Specially, two types of chimera states are discovered. For chimera
states at weak coupling strength, FHN units oscillate near the
equilibrium $(u^*,v^*)$. On the other hand, chimera states at
strong coupling strength display strong characteristics of the
excitability.

The chimera state at weak coupling strength $\sigma=0.05$ is shown
in Fig.~\ref{fig1}.  The snapshot of the activator variables $u_i$
in Fig.~\ref{fig1}(a) suggests a chimera state with four same size
coherent clusters, separated by four same size narrow incoherent
clusters. The snapshot of the phases of FHN units, defined as
$\Theta_i=\arctan(\dot{v}_i/\dot{u}_i)$, in Fig.~\ref{fig1}(b)
shows that adjacent coherent clusters are in antiphase, which is
same as those clustered chimera states observed in nonlocally
coupled phase oscillators \cite{set08,zhu12}. (To be noted, in the
case of coexisting large- and small-amplitude oscillations in Fig.~\ref{fig2}, the definition of the phase $\Theta_i$
by $\dot{v}_i$ and $\dot{u}_i$ is much more convenient than the
one by $v_i$ and $u_i$ in calculating the mean phase velocities of
FHN units). The coherent and incoherent clusters can be
distinguished from each other through the mean phase velocity of
each FHN unit, defined as
$\omega_i=\langle\dot{\Theta}_i\rangle_t$ with
$\langle\cdot\rangle_t$ the time average over a long time
interval. The profile of $\omega_i$ in Fig.~\ref{fig1}(c) shows
that FHN units in the coherent clusters have the same mean phase
velocity, denoted as $\Omega$, while those in the incoherent
clusters have different $\omega_i$ which are always larger than
$\Omega$. The snapshots of FHN units in the $(u_i,v_i)$ plane at
different specific times in Fig.~\ref{fig1}(d1)-(d3) show that the
units in coherent clusters are concentrated on two pieces of
curves opposite to the equilibrium and units from adjacent
coherent clusters fall into different curves. In contrast, FHN
units in incoherent clusters are scattered on noisy orbits between
these two curves.

\begin{figure*}
\includegraphics[width=6in]{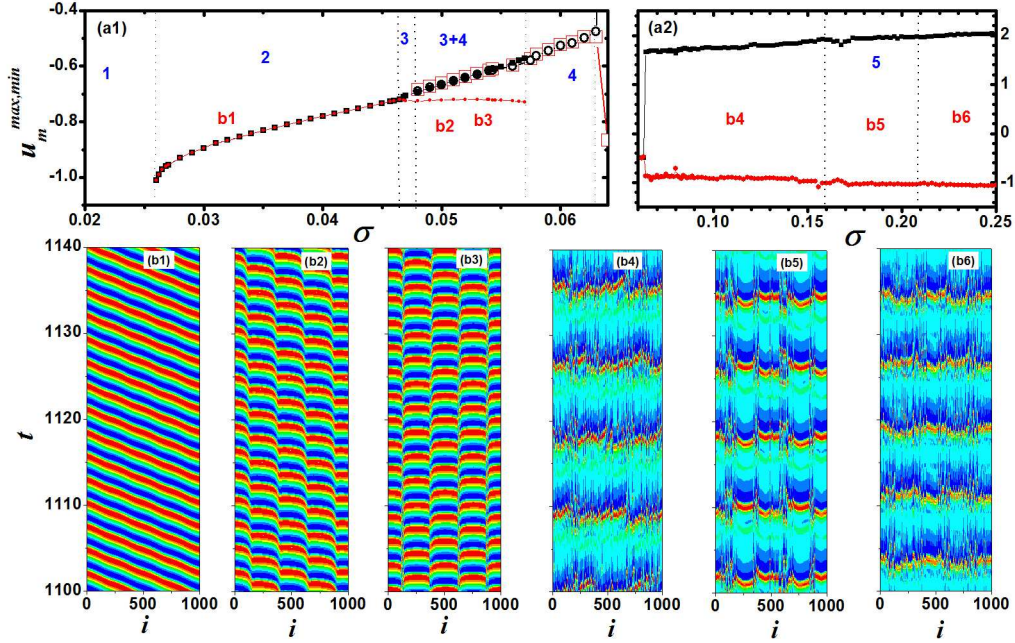}
\caption{\label{fig3}(color online)(a1) and (a2) $u_m^{max}$ (in
black) and $u_m^{min}$ (in red) are plotted against $\sigma$.
Solid symbols are obtained from the forward continuation and open
symbols from the backward continuation. The numbers denote
different dynamical regimes (see main text). In the regime denoted
by $"3+4"$, the modulated travelling wave states and the
phase-chimera states coexist. (b1)-(b6) The spatial-temporal plots
of $u_i$ at the coupling strengthes denoted by b1-b6 in (a).}
\end{figure*}

The FHN units in coherent and incoherent clusters obey different
dynamics. Figure~\ref{fig1}(e) shows the trajectories of two
typical FHN units, the limit cycle for the coherent unit and the
torus-like trajectory for the incoherent unit. The torus-like
trajectory is enclosed by the limit cycle and both trajectories
are localized around the equilibrium $(u^*,v^*)$. $u(t)$ of the
two units in Fig.~\ref{fig1}(f) show that the incoherent unit
oscillates faster than the coherent one and its amplitude varies
with time. Furthermore, the evolutions of the maximum of $u(t)$,
$u_m(t)$, for the two units in Figs.~\ref{fig1}(g) and (h) show a
periodic oscillation for the coherent unit while irregular
dynamics for the incoherent unit.

As the coupling strength $\sigma$ is increased into a strong
couple strength region, an different and interesting type of
chimera state is emerged and presented in Fig.~\ref{fig2} at
coupling strength $\sigma=0.18$. In comparison with the one in
Fig.~\ref{fig1}, there are two remarkable differences. Firstly,
the profile of $\omega_i$ in Fig.~\ref{fig2}(c) indicates that
coherent FHN units oscillate faster than incoherent FHN units.
Secondly, the snapshots of FHN units in the $(u_i,v_i)$ plane in
Fig.~\ref{fig2}(d1)-(d3) remarkably show that the oscillation of
FHN units is not localized near $(u^*,v^*)$. Instead, FHN units
shift back and forth between the low amplitude oscillation around
the equilibrium and the high amplitude oscillation with the
excursion characteristics of excitable dynamics. It can be further
evidenced by the trajectories of one coherent unit in
Fig.~\ref{fig2}(e) and one incoherent unit in Fig.~\ref{fig2}(f).
The time sequence of $u$ in Fig.~\ref{fig2}(g) shows that the
coherent unit alternates between two intertwined temporal
patterns, one cycle of small amplitude oscillation followed by one
cycle of high amplitude oscillation, and two cycles of small
amplitude oscillation followed by one cycle of high amplitude
oscillation. In contrast, the incoherent unit displays irregular
behaviors between the low and the high amplitude oscillations in
Fig.~\ref{fig2}(h). In addition, the $\pi$ phase difference
between the adjacent coherent clusters is not held any more, which
is clearly shown in Fig.~\ref{fig2}(b).

\section{The emergence of collective oscillation and the transition to chimera dynamics}
How can FHN units support chimera states when they are in
excitable regime? In other words, where does the oscillation for
each unit emerge from? Our work reveals that a collective motion
can be induced through the nonlocal coupling. If the coupling
strength is higher than a critical value, the homogeneous
equilibrium $(u^*,v^*)$ becomes unstable and a collective motion
in the form of travelling wave emerges. To elucidate it, we
analyze the stability of the homogeneous equilibrium $(u^*,v^*)$
in details. The evolution of the perturbation $(\delta u_i,\delta
v_i)$ near $(u^*,v^*)$ follows
\begin{eqnarray}\label{eq3}
\frac{d}{dt}(\begin{array}{c}
              \delta u_i \\
              \delta v_i
            \end{array})=DF(u^*,v^*)(\begin{array}{c}
              \delta u_i \\
              \delta v_i
            \end{array})+\Sigma_{j=i-R}^{i+R}\bar{B}(\begin{array}{c}
              \delta u_j-\delta u_i \\
              \delta v_j-\delta v_i
            \end{array})
\end{eqnarray}
with $DF(u^*,v^*)=\left(
                    \begin{array}{cc}
                      (1-u^{*2})/\epsilon & -1/\epsilon \\
                      1 & 0 \\
                    \end{array}
                  \right)$ and $\bar{B}=\left(
    \begin{array}{cc}
      b_{uu}/\epsilon & b_{uv}/\epsilon \\
      b_{vu} & b_{vv} \\
    \end{array}
 \right)$. For the model (1), there are $N$ spatial modes characterized by
the vectors $\nu_k=(1,\omega_k,\omega_k,...,\omega_k^{N-1})^T/\sqrt{N}$,
($k=1,2,...,N-1$) where $\omega_k=\exp(\mathbf{i}2\pi k/N)$ with
$\mathbf{i}$ the imaginary unit and $k$ the wave number. Expanding
the perturbation over these spatial modes and substituting it into
the Eq.~(\ref{eq3}), we have
\begin{eqnarray}\label{eq4}
\dot{\delta}_k=(DF(u^*,v^*)+\sigma\lambda_k \bar{B})\delta_k=DF^{(k)}(u^*,v^*)\delta_k
\end{eqnarray}
with
\begin{eqnarray}\label{eq5}
\lambda_k&=&-1+\frac{\cos\frac{2\pi kR}{N}-\cos\frac{2\pi k(R+1)}{N}+\cos\frac{2\pi k}{N}-1}{2R(1-\cos\frac{2\pi k}{N})}
\end{eqnarray}
Either the determinant $Det(DF^{(k)})<0$ or the trace
$Tr(DF^{(k)})>0$ when $Det(DF^{(k)})>0$ leads the spatial mode
$\nu_k$ to be unstable. The latter leads $\delta_k$ to grow in an
oscillating way and the former leads $\delta_k$ to grow
monotonically in an exponential way. It can be shown that
$Det(DF^{(k)})>0$ is always satisfied for the parameters used in
this letter. Therefore, once $Tr(DF^{(k)})>0$, the spatial mode
$\nu_k$ becomes unstable through a Hopf bifurcation. For each
spatial mode, we have its critical coupling strength
$\sigma^{(k)}_c=(a^2-1)/(1+\epsilon)\lambda_k \cos\phi$. Since
$\lambda_k$ is always negative, $\cos\phi<0$ is required to have a
positive $\sigma^{(k)}_c$. Clearly, the most unstable spatial mode
$\nu_{k^{*}}$ has the largest $|\lambda_k^*|$. When
$\sigma>\sigma^{(k^{*})}_c$, the homogeneous equilibrium becomes
unstable and a travelling wave with the wave number $k^{*}$
appears and renders the oscillation to each FHN unit. For the
parameters in Figs.~\ref{fig1} and \ref{fig2}, we obtain $k^{*}=2$
and $\sigma^{(k^{*})}_c=0.026$. The low amplitude oscillation
there roots in the collective oscillation while the high amplitude
oscillation results from the limit cycle induced by the
excitability of local units. In this respect, we call the chimera
states in Fig.~\ref{fig1} the phase-chimera states and the ones in
Fig.~\ref{fig2} the excitability-chimera states.

\begin{figure}
\includegraphics[width=4.5in]{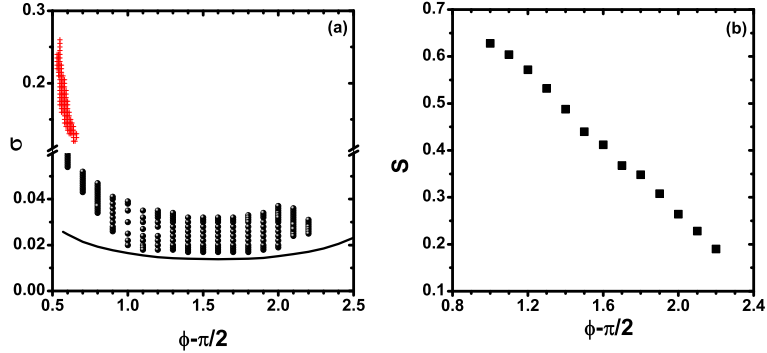}
\caption{\label{fig4}(color online)(a) The stability regimes for
phase-chimera states (black dots) and excitability-chimera states
(red plus) in the plane of $\sigma$ and $\phi$. The black curve denotes
the onset of the collective motion below which the homogeneous equilibrium is stable. (b) The
dependence of the fraction of coherent units, $S$, on $\phi$, at
$\sigma=0.03$.}
\end{figure}

\begin{figure}
\includegraphics[width=4.5in]{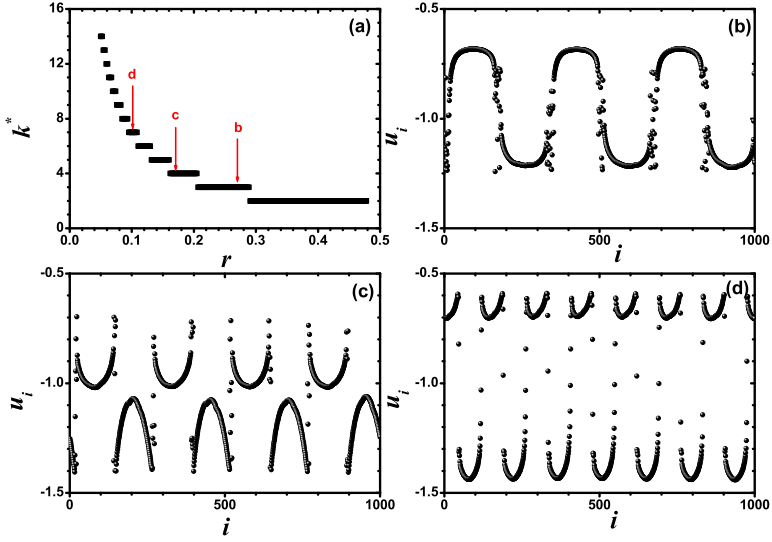}
\caption{\label{fig5}(color online)(a) The wave number $k^*$ of
the most unstable spatial mode to the homogeneous equilibrium
$(u^*,v^*)$ is plotted against $r$. Red alphabets b, c, and d
denote parameter values for plots (b) , (c), and (d). The
snapshots of the variables $u_i$ at $r=0.27$ and $\sigma=0.05$ in
(b), at $r=0.17$ and $\sigma=0.06$ in (c), at $r=0.1$, and
$\sigma=0.06$ in (d).}
\end{figure}

To study the transition scenario from the travelling waves to the
chimera states, we consider two types of bifurcation diagrams,
forward continuation and backward continuation. The coupling
strength $\sigma$ is successively increased (or decreased) by a
$\delta\sigma$ in the forward (or backward) continuation and the
initial conditions for one coupling strength are the final state
of the previous one. We monitor the evolution of $u_m$ for a
coherent FHN unit (or for an arbitrary FHN unit if there is no
coherent one). The maximum and the minimum of $u_{m}$, $u_m^{max}$
and $u_m^{min}$, are measured and their dependence on $\sigma$ are
presented in Fig.~\ref{fig3}(a1) and (a2). There are several
dynamical regimes. The homogeneous equilibrium $(u^*,v^*)$ is
stable for $\sigma<0.026$ (the regime 1). In the regime 2 where
$u_m^{max}=u_m^{min}$, the coherent unit oscillates periodically
and its oscillation amplitude $u_m^{max}-u^*$ increases with
$\sigma$. As shown by the spatial-temporal plot of $u$ in
Fig.~\ref{fig3}(b1), travelling wave states are realized in this
regime. Increasing $\sigma$ from the regime 2, the travelling wave
state becomes unstable at $\sigma\approx0.046$ and a modulated
travelling wave (see Fig.~\ref{fig3}(b2)) occurs in the regime 3.
$u_m^{max}\neq u_m^{min}$ in the regime 3, which suggests the
quasiperiodic motion for FHN units. At stronger coupling strength
$\sigma$, $u_m^{max}$ becomes equal to $u_m^{min}$ again as shown
in the regime 4 in which the phase-chimera state is realized (see
Fig.~\ref{fig3}(b3)). By comparing the results from the forward
and backward continuations, we find that the phase-chimera states
coexist with the modulated travelling waves in the range
$0.048<\sigma<0.057$. The phase-chimera state becomes unstable
around $\sigma=0.063$ at which the oscillation amplitudes of FHN
units become strong enough to trigger the excitability of FHN
units. Consequently, the oscillations of FHN units are not
localized near their equilibria as $\sigma$ is beyond $0.063$.
They always shift back and forth between the small amplitude and
high amplitude oscillations characterized by $u_m^{min}$ and
$u_m^{max}$ respectively. Together with Figs.~\ref{fig3}(a2), the
spatial-temporal plots of $u$ in Figs.~\ref{fig3}(b4) -(b6)
indicate that excitability-chimera state only exists in the regime
5 and there is no clear difference between the dynamics in the
regimes 4 and 6.

\section{The impact of the model parameters on chimera dynamics}
To further investigate the chimera dynamics, we present in Fig.~\ref{fig4}(a) the phase
diagram in the plane of $\sigma$ and $\phi$ in which the regime
with red pluses supports the excitability-chimera is denot while
the regime with black circles supports the phase-chimera. As shown
in Fig.~\ref{fig4}(a), the excitability-chimera
states exist within a narrow range of $\phi$ at strong coupling
strength. On the other hand, the phase-chimera states can exist in
the parameter regime with $\phi$ spanning from 2.1 to 3.8 but a
narrow range of $\sigma$. The parameter $\phi$ has strong impacts
on the fraction of the coherent units in the population, defined
as $S=N_c/N$ with $N_c$ the number of coherent units, for the
phase-chimera state. Figure~\ref{fig4}(b) clearly shows a linear
dependence of $S$ on $\phi$. We also numerically find that the
total number of coherent clusters in the phase-chimera state is
twice of  the wave number $k^*$, which is mainly determined by the
coupling radius $r$. The relation of $k^*$ against $r$ is shown in
Fig.~\ref{fig5}(a). It clearly shows that $k^*$ tends to 2 even
when $r$ approaches 0.5. This suggests nonexistence of the
phase-chimera state with 1 or 2 coherent clusters in the
nonlocally coupled FHN system (1). The snapshots of the variable
$u_i$ at three different $r$ are shown in Fig.~\ref{fig5}(b)-(d),
respectively. The corresponding number of coherent clusters is 6,
8 and 14, which is exact twice of the wave number $k^*$.

\section{Conclusions}
In previous investigations on chimera dynamics,
the isolated unit has to be oscillatory, either self-oscillation
or noise-induced-oscillation due to the coherent resonance. In
this paper, we reported the existence of chimera states in a
nonlocally coupled FHN system in the excitable regime where
isolated unit only allows for the solution of equilibrium. The
findings in this paper suggest that the emergence of chimera
states is not restricted to oscillatory systems. We find that the
collective oscillation induced by nonlocal coupling plays critical
roles for the emergence of chimera states in this system. For
sufficiently weak coupling strength, the nonlocally coupled
systems only support homogeneous equilibrium. However, when the
coupling strength is beyond a critical value, the homogeneous
equilibrium becomes unstable and a travelling wave state is born,
which makes units to oscillate and makes the appearance of chimera
states to be possible. Depending on the coupling strength, two
types of chimera states are observed. The first type of chimera
state occurs at weak coupling strength in which both coherent and
incoherent FHN units oscillate with low amplitude around the
equilibrium of isolated unit. The second type of chimera state
appears in the strong coupling strength, in which the dynamics of
both coherent and incoherent units shifts back and forth between
the low amplitude oscillation around their equilibria and the high
amplitude oscillation following the excited limit cycle. We also
numerically find that the total number of coherent clusters in
these chimera states is determined by the coupling radius and is
twice of the wave number of the travelling wave born out of the
homogeneous equilibrium. Although our revealed findings are based
on the FHN systems, the emergence of chimera states by nonlocal
coupling induced collective oscillation should not depend on the
specific system, and can be realized in other excitable systems.

\section*{Acknowledgements}
This work was supported by National Natural Science Foundation of China (Grant
Nos. 11575036 and 11505016).

\end{document}